\documentclass[10pt,a4paper]{article}
\usepackage[utf8]{inputenc}
\usepackage[english]{babel}
\usepackage{graphpap,amsmath,amsfonts,amssymb,epsfig,makeidx,graphicx,textcomp,cite,citeall,authblk,subfloat,float,xcolor} 
\usepackage[left=1.5cm, right=1.5cm, top=1.5cm, bottom=1.5cm]{geometry}

\author[1]{{Soham Chandra} \thanks{E-mail addresses: soham.rs@presiuniv.ac.in ; sohamc07@gmail.com}}

\affil[1]{\textit{\normalsize{Department of Physics, Presidency University, 86/1 College Street, Kolkata -700 073, India}}}

\vspace{20pt}

\title{\textbf{A Monte Carlo study on the temperature dependence of hysteresis loops in Ising Spin-1 Square Bilayers}}

\date{} 
\begin{document}
	\maketitle	
	\begin{abstract}
		A Metropolis Monte Carlo simulation is used in this paper to investigate the temperature dependency of the hysteresis loops of a spin-1 bilayer with \textit{square monolayers}. In this system, the atoms interact ferromagnetically in-plane, with either ferromagnetic or antiferromagnetic interplane interactions. The effects of four distinct combinations of the Hamiltonian parameters on the hysteresis behaviours are discussed in detail. The geometry of the hysteresis loops changes depending on how the exchange couplings are combined. With ferromagnetic interlayer coupling, only the central hysteresis loop opens while for the antiferromagnetic case, the hysteresis loop becomes a double loop for the specific combination of coupling strengths. Additionally, in all these cases, the area of the hysteresis loops grows with the gradual lowering of the temperature. 
	\end{abstract}

\vskip 2cm
\textbf{Keywords:} Spin-1 Ising square bilayer; Metropolis Monte Carlo simulation; Hysteresis loops 

\twocolumn
\section{Introduction}
\label{sec_intro}

Layered magnetic materials often show very different physical properties than the bulk. The experimental interest in layered magnets has rapidly increased since the discovery of thin-film growth techniques, e.g., metalorganic chemical vapor deposition (MOCVD) \cite{Stringfellow}, molecular-beam epitaxy (MBE) \cite{Herman}, pulsed laser deposition (PLD) \cite{Singh}, atomic layer deposition (ALD) \cite{George}. The growth of bilayered \cite{Stier}, trilayered \cite{Leiner}, and multilayered \cite{Sankowski,Maitra} systems with desired properties have become a reality thanks to such experimental breakthroughs. Expectedly, theoretical and computational studies of layered magnets have also gained momentum. In multilayered magnets, the magnetizations of each of the layers may evolve differently with temperature, depending upon the coupling strengths. Many fascinating \textit{bulk} phenomena, such as \textit{compensation} \cite{Diaz,Chandra,Chandra2}, result from the combined effect of such various magnetic behaviours. In equilibrium statistical physics, thin films have been investigated using Ising mechanics by various established theoretical methods e.g. mean-field theory (MFT) \cite{Mills,Lubensky,Hong}, spin-fluctuation theory \cite{Benneman}, the effective field theory (EFT) \cite{Kaneyoshi,Akinci}, the series-expansion method \cite{Oitmaa}, the exact recursion equation on the Bethe lattice \cite{Albayrak}, the cluster variation method in pair approximation \cite{Balcerzak}, the renormalization-group (RG) method \cite{Ohno,Jiang}, and Monte Carlo (MC) simulations \cite{Moussa,Laosiritaworn,Chou,Candia,Albano}.

The hysteresis behaviour of a Spin-$1$ Ising bilayer with just nearest neighbour magnetic interactions is the subject of this article. Different types of hysteretic characteristics are required for various magnetic applications. Magnetic materials for stable information storage, for example, must be strongly hysteretic, but generator magnets must have minimal hysteresis to minimise energy losses \cite{Tebble}. Recently, photonic metasurfaces and coupled nonlinear metasurfaces \cite{Costas1,Costas2} are being investigated in the context of multistability. To achieve the desired characteristics, popular optimisation and tuning techniques of magnetic materials include different chemical and structural means e.g. composition and segregation of compounds, synthesis of micro and nanostructures etc \cite{Chen}. But even these highly successful methods have a very common drawback. After a material is prepared, the physical properties become fixed and thus requires changes in the structural or compositional level to achieve different physical properties. So it is desirable to predict the physical behaviour beforehand and Monte Carlo simulation can be an answer to that.

Let us discuss a few experimental and theoretical advances in the past decades on magnetic bilayers. In \cite{Berger1}, the authors have experimentally realized a magnetic bilayer which consists of a hard-magnetic $CoPtCrB$-layer (HL) as the tuning layer which is antiferromagnetically coupled to a tunable soft-magnetic $Co$ layer (SL) by a $6$\AA{} thick $Ru$ film, in which the magnetic hysteresis loop properties can be tuned. This tuning procedure has been shown in \cite{Berger2}, to be temperature independent and reversible, besides being highly effective. Ferromagnetic (FM)/ antiferromagnetic (AFM) bilayers are popularly investigated for exchange bias (EB) \cite{Meiklejohn,Berkowitz-Nogues-Kiwi} which is important for magnetoelectronic devices \cite{Childress}. Depending upon the values of the cooling field, in a $Gd_{40}Fe_{60}/Tb_{12}Fe_{88}$ exchange-coupled bilayer system, both negative and positive EB are observed \cite{Mangin}. In \cite{Dumesnil}, strong temperature dependence of magnetization reversal is shown in a  $[DyFe_{2} (5 nm)/$ $YFe_{2} (20 nm)]$ superlattice, antiferromagnetically coupled at their interfaces by recording compound-specific hysteresis loops. From the theoretical perspective, Effective field theory (EFT) with correlations has been applied to various similar kind of systems \cite{Kaneyoshi2}. The Compensation and hysteresis behaviours of a spin-1 bilayer Ising model are studied by effective field theory with correlations in \cite{Kantar}. On such a system, thus it is desirable to perform a Monte Carlo simulation to get a detailed idea of the microscopic behaviour. Thus in the current study, it will be shown by extensive Monte Carlo simulations, how the shape and the area of hysteresis loops of the spin-$1$ Ising bilayer change in response to the change in temperature and combination of exchange coupling strengths. At this point, we should recall that spin-$1$ anisotropic nearest neighbour interacting systems are the simplest systems where the effect of non-zero crystalline anisotropy is observed on the critical behaviour in model magnetic systems (e.g. Blume-Capel Model). But in this study, we would restrict ourselves to the Ising case (absence of any crystalline anisotropy).
 
\begin{figure}[!htb]
	
		\resizebox{9.5cm}{!}{\includegraphics[angle=0]{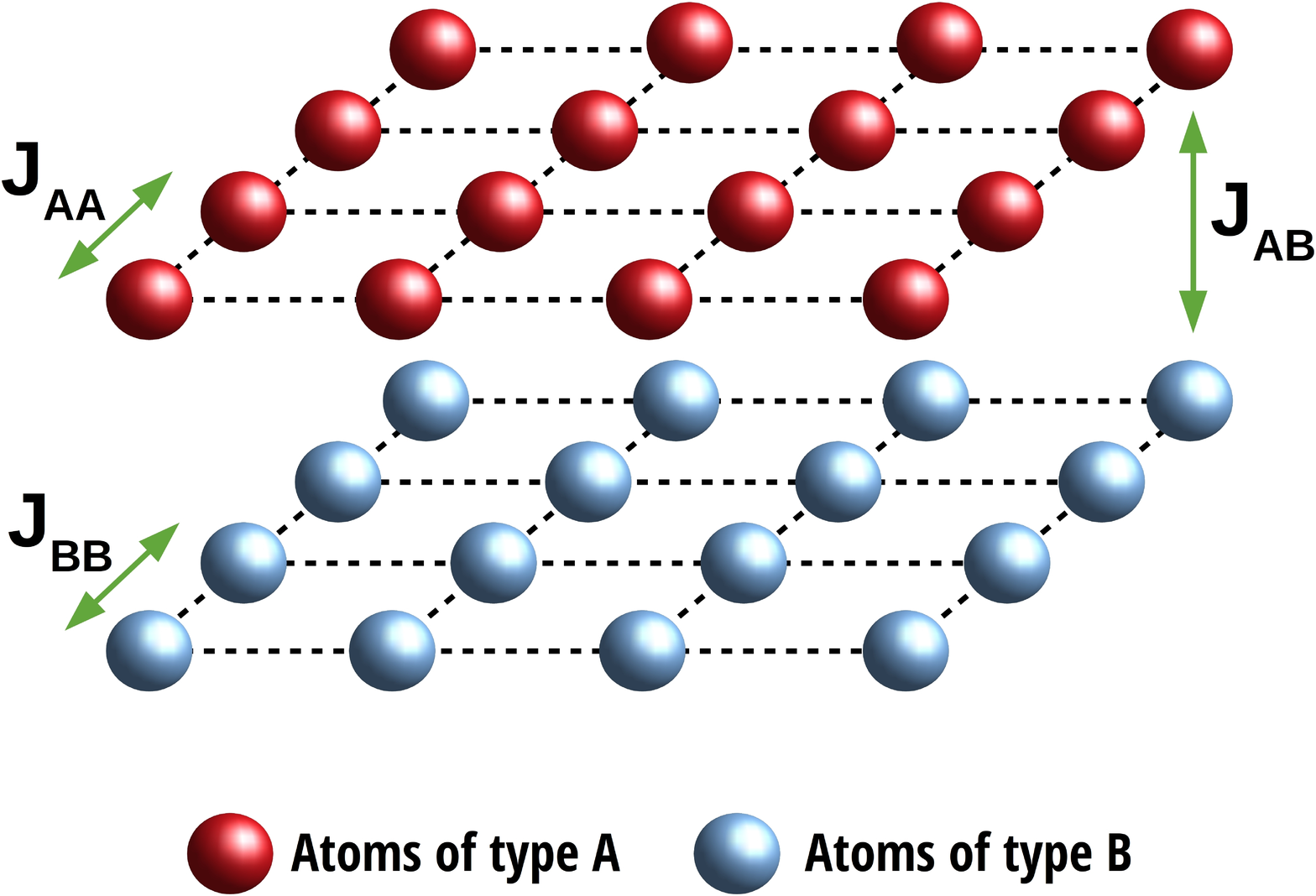}}
		
		\caption{ (Colour Online) Miniaturised versions ($2\times4\times4$) of a AB-type square bilayered Ising superlattice, $A$ and $B$. Each of the sublattices of the ferrimagnetic systems are formed on square lattice. The actual simulation is carried out on a system with $N_{sites}=2\times100\times100$ .} 
		
		\label{fig_1_lattice_structure}
\end{figure}

Figure \ref{fig_1_lattice_structure} shows the system employed in this investigation. The top layer consists of type-A theoretical atoms, whereas the bottom layer consists of B-type atoms. In-plane and inter-plane, the spins interact Ising-like between nearest neighbours. Let us assume that, $sigma$ and $S$ are the spin variables for the A and B-atoms (of spin-$1$), respectively. Both $\sigma^{z}$ and $S^{z}$ can have projections $0$,$\pm 1$. As a result, the Hamiltonian for such a bilayered Ising superlattice is:
\begin{eqnarray}
\nonumber
H = &-& 
J_{AA}\sum_{\langle i,j \rangle}\sigma_{i}^{z}\sigma_{j}^{z} - J_{BB}\sum_{\langle m,n \rangle}S_{m}^{z}S_{n}^{z} \\
\label{eq_Hamiltonian}
& - &
J_{AB} \sum_{\langle i,m \rangle}\sigma_{i}^{z}S_{m}^{z} -  h(\sum_{i}\sigma_{i}^{z}+\sum_{m}S_{m}^{z})
\end{eqnarray}
$\langle i,j \rangle$, $\langle m,n \rangle$ are nearest-neighbor pairs in the top A-layer and bottom B-layer respectively and $\langle i,m\rangle$ nearest-neighbor inter-plane sites. The first two terms are intra-planar ferromagnetic ones. The third term is for the inter-planar nearest neighbour interactions, between the layers. The fourth term denotes the \textit{spin-field interaction term} of all the spins to the external magnetic field of intensity, $h$. To represent ferromagnetic interactions: $J_{AA}>0$ and $J_{BB}>0$. For ferromagnetic inter-layer interactions, $J_{AB}>0$ and for anti-ferromagnetic inter-layer interactions, $J_{AB}<0$.

The remainder of the paper is organised as follows. In Section \ref{sec_simulation}, the simulational details are described. Section \ref{sec_results} contains the numerical results and associated discussions. The final conclusions of the paper are offered in Section \ref{sec_conclusion}.

\section{Simulation Protocol}
\label{sec_simulation}

The Metropolis single spin-flip algorithm \cite{Metropolis} is employed for simulation of the system \cite{Landau, Newman, Binder} of Figure \ref{fig_1_lattice_structure}. Both the monolayers are formed on a square lattice with $L^{2}$ sites where we have taken $L=100$. Only the $z$-components of spin projections of nearest neighbours, $\sigma_{i}^{z}$ and $S_{m}^{z}$ $(\sigma_{i}^{z}=0,\pm 1; S_{m}^{z}=0,\pm 1)$ contribute to the cooperative and spin-field interactions. For a few cases, increasing the lattice size from $L=100$ has been found to have no detectable effect on the obtained results previously. So the chosen lattice size is statistically reliable for simulation. The system is initiated at a high-temperature paramagnetic phase, with randomly selected one-third of the total spin projections in the top A-layer being in $\sigma_{i}^{z}=+1$, another one-third being in $\sigma_{i}^{z}=0$ and the other one-third being in $\sigma_{i}^{z}=-1$. The bottom B-layer is also similarly initiated. At a fixed temperature $T$, the dynamics of spin flipping obeys the Metropolis rate \cite{Metropolis, Newman}, of Equation [\ref{eq_metropolis}] (for reference, written in terms of the spins $S_{i}$):
\begin{equation}
\label{eq_metropolis}
P(S_{i,initial}^{z} \to S_{i,final}^{z}) = \text{min} \{1, \exp (-\Delta E/k_{B}T)\}
\end{equation}
Here $\Delta E$ denotes the associated change in the energy in changing the $i$-th spin projection from $S_{i}^{z} (\sigma_{i}^{z})$ to any of the three available spin orientations. One Monte Carlo sweep (MCS) of the entire system is defined by such $2L^{2}$ random, single-spin updates. In this work, \textit{one MCS} defines the unit of time. \textit{Periodic boundary conditions in-plane and Open boundary conditions along the vertical are employed.}

The system spends $10^{5}$ MCS at every temperature step. The last configuration at the previous \textit{higher} temperature acts as the starting configuration at a new \textit{lower} temperature. In a field-free environment, for \textit{equilibration} we would discard the first $5\times10^{4}$ MCS. This much time is checked to be sufficient to reach equilibrium. For the next $N=5\times10^{4}$ MCS, we would collect data for relevant physical quantities (e.g. magnetization and cooperative energy). The temperatures are measured in units of $J_{BB}/k_{B}$. The intra-planar ferromagnetic coupling strength $J_{BB}$ is chosen to be the dominant one and set to $1.0$. The intra-planar ferromagnetic ratio, $J_{AA}/J_{BB}$ is also fixed at $1.0$. The inter-planar coupling ratio $J_{AB}/J_{BB}$ is either $0.50$ (ferromagnetic) or $-0.50$ (anti-ferromagnetic). For any combination of $J_{AA}/J_{BB}$ and $J_{AB}/J_{BB}$, the time/ ensemble averages of the following quantities are calculated after equilibration at any temperature, $(T)$ in the following manner \cite{Diaz,Chandra}:\\
\textbf{(1) Sublattice magnetisations} are calculated after the $j$-th MCS by:
\begin{equation}
M_{q,j}(T)=\frac{1}{L^{2}}\sum_{x,y=1}^{L} \left( S_{q,j}^{z}(T)\right)_{xy}
\end{equation}
The time/ensemble averaged sublattice magnetizations are calculated by:
\begin{equation}
\langle M_{q}(T)\rangle =\frac{1}{N}\sum_{j=1}^{N} M_{q,j}(T)
\end{equation}
where $q$ is to be replaced by $t\text{ or }b$ for top and bottom layers.\\
\textbf{(2) The total magnetization} serves as \textit{the order parameter}, $M(T)$, for the Ising bilayer at temperature, $T$ and is defined as:
\begin{equation}
M(T)=\frac{1}{2}(\langle M_{t}(T)\rangle+\langle M_{b}(T)\rangle)
\end{equation}
\\
\textbf{(3) Fluctuation of the order parameter,} $\Delta M(T)$ at temperature, $T$, is found out as follows:
\begin{equation}
{\Delta M}(T)=\sqrt{\frac{1}{N} \sum_{j=1}^{N} \left[M_{j}(T)-M(T)\right]^{2} }
\end{equation}
where $M_{j}(T)$ is the total magnetisation of the whole system calculated at the (\textit{j-th MCS}), at temperature, $T$, by:
\begin{equation}
M_{j}(T)=\frac{1}{2}( M_{t,j}(T)+ M_{b,j}(T))
\end{equation} 

\textbf{Fluctuation of the cooperative energy per site}, $\Delta E(T)$ at temperature, $T$, is found out in a similar manner:
\begin{equation}
{\Delta E}(T)=\sqrt{\frac{1}{N}\sum_{j=1}^{N} \left[E_{j}(T)-E(T)\right]^{2} }
\end{equation}
where $E_{j}(T)$ is the cooperative energy per site calculated at the \textit{j-th MCS}, at temperature, $T$ and $E(T)$ is the associated ensemble average. $E_{j}(T)$ is calculated from the Equation \ref{eq_Hamiltonian}, after setting $h=0$.

At the \textit{pseudo-critical} temperatures, the fluctuations peak. Any refinement of the values of the critical point is not required in this study as we would be observing hysteresis well below the critical temperatures. So an indication of the value of the critical temperature is just enough for our purpose. While cooling, when we land on a pre-selected temperature for hysteresis, we would first equilibrate the system. A constant, high and positive magnetic field will then be switched ON and we would allow $10^{4}$ MCS for the transients to die. For the next $2\times 10^{4}$ MCS, we would collect the data for the magnetization of the system. The external magnetic field will be lowered in small steps and for each value of it, we would repeat the above procedure. We would stop at the negative of the initial field to complete the Forward/Descending loop. The change of magnetic field and the associated magnetic response in the reverse order in a similar fashion is called the Reverse/Ascending loop. In this study we have, $h\in[-5,+5]$ with $\Delta h=0.10$. The Jackknife method \cite{Newman} is used to provide an estimate of the errors with the magnetizations and fluctuations . 
\section{Results and Discussion}
\label{sec_results}
In this section, we study the hysteresis behaviour of the spin-1 bilayer Ising model as outlined in Section \ref{sec_simulation}. For two particular choices of the Hamiltonian parameters of the system, we would see how it affects the hysteresis loops.
\subsection{Interlayer coupling: Ferromagnetic}
\label{subsec_inter ferro}
\begin{figure*}[!htb]
	\begin{center}
		\begin{tabular}{c}
			
			\resizebox{7.5cm}{!}{\includegraphics[angle=0]{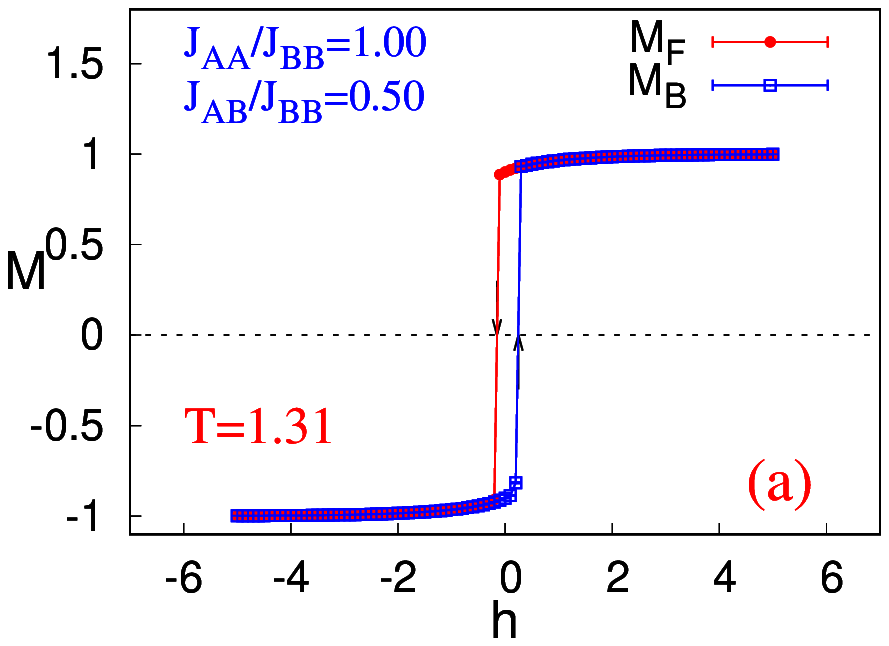}}
			\resizebox{7.5cm}{!}{\includegraphics[angle=0]{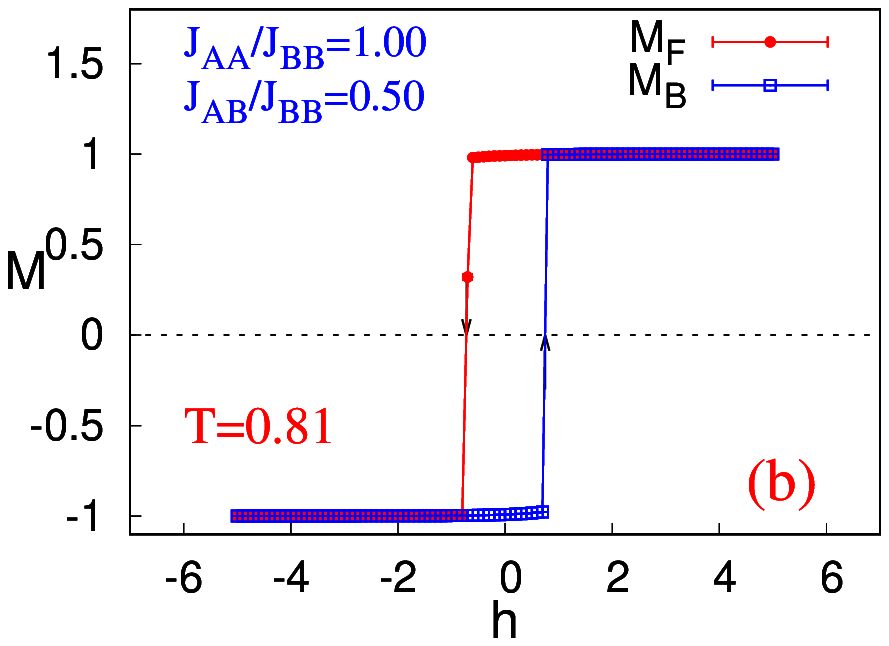}}\\
			
			\resizebox{7.5cm}{!}{\includegraphics[angle=0]{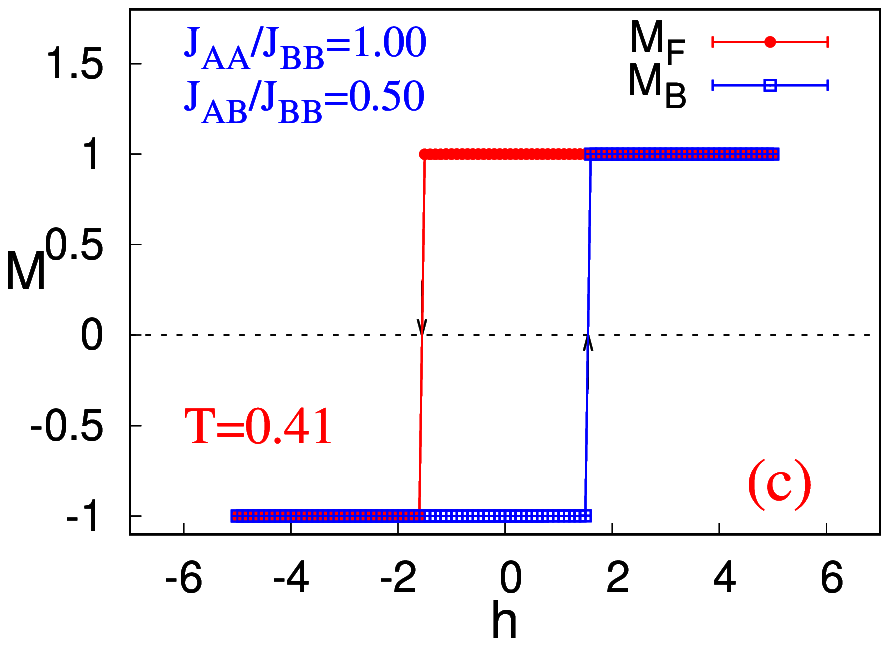}}
			\resizebox{7.5cm}{!}{\includegraphics[angle=0]{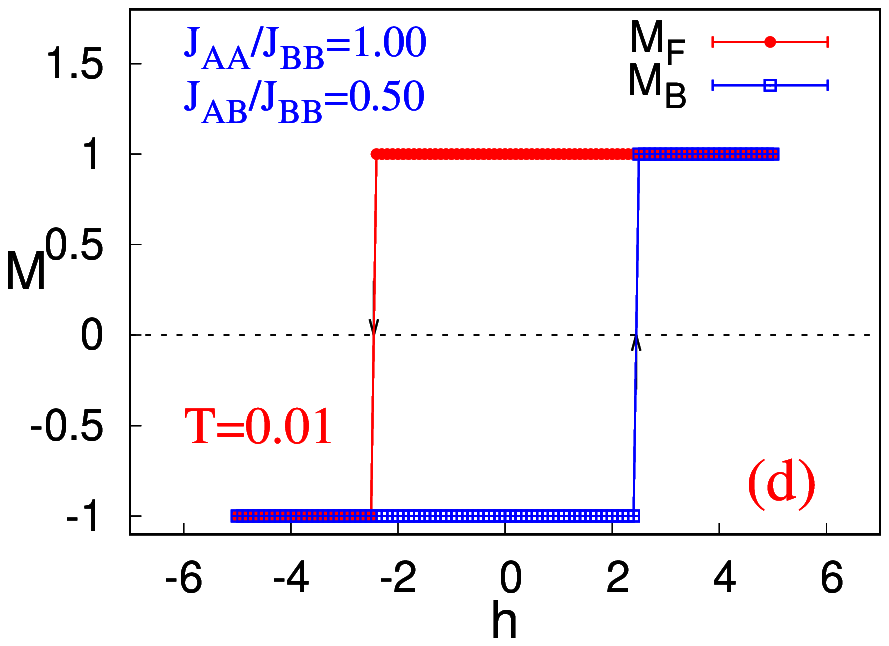}}
			
		\end{tabular}
		\caption{ (Colour Online) Plots of total magnetization versus the applied magnetic field. The effect of the temperature on the hysteresis behaviour for $T=1.31,0.81,0.41,0.01$ with the fixed values of $J_{AA}/J_{BB}=1.00$ and $J_{AA}/J_{BB}=0.50$. The descending/forward (RED, magnetisation $M_{F}$) and ascending/backward (BLUE, magnetisation $M_{B}$) external field directions are represented by arrows on each of the magnetization curves. As temperature of the system is gradually lowered, the area of the hysteresis loops increases.}
		\label{fig_2_hys_fr_fr}
	\end{center}
\end{figure*}

\begin{figure*}[!htb]
	\begin{center}
		\begin{tabular}{c}
			
			\resizebox{7.5cm}{!}{\includegraphics[angle=0]{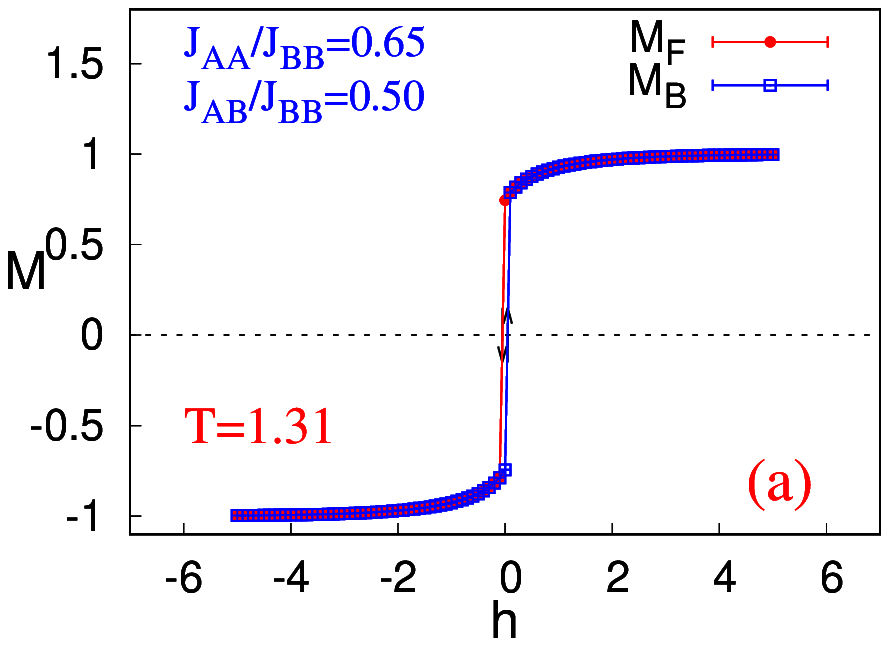}}
			\resizebox{7.5cm}{!}{\includegraphics[angle=0]{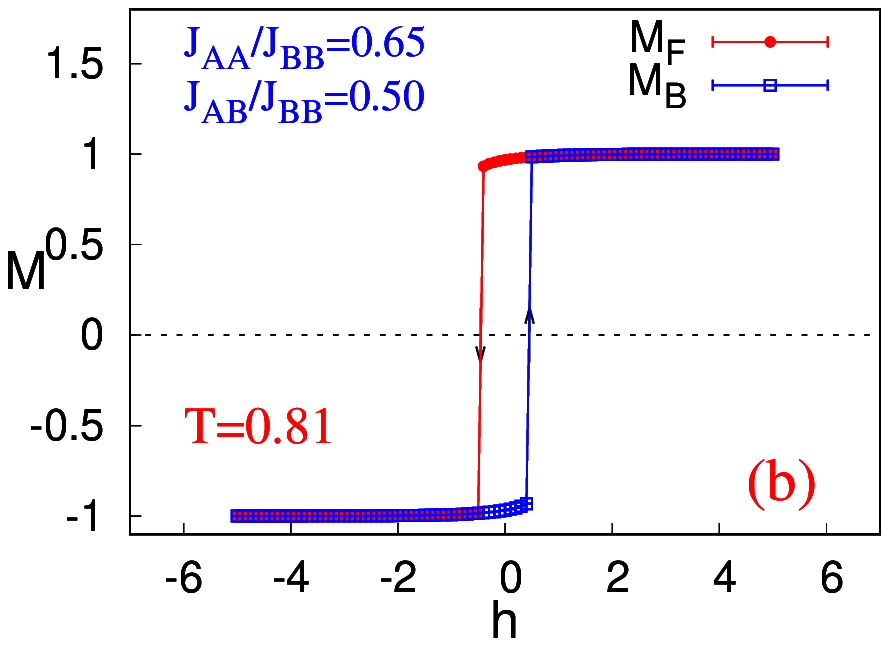}}\\
			
			\resizebox{7.5cm}{!}{\includegraphics[angle=0]{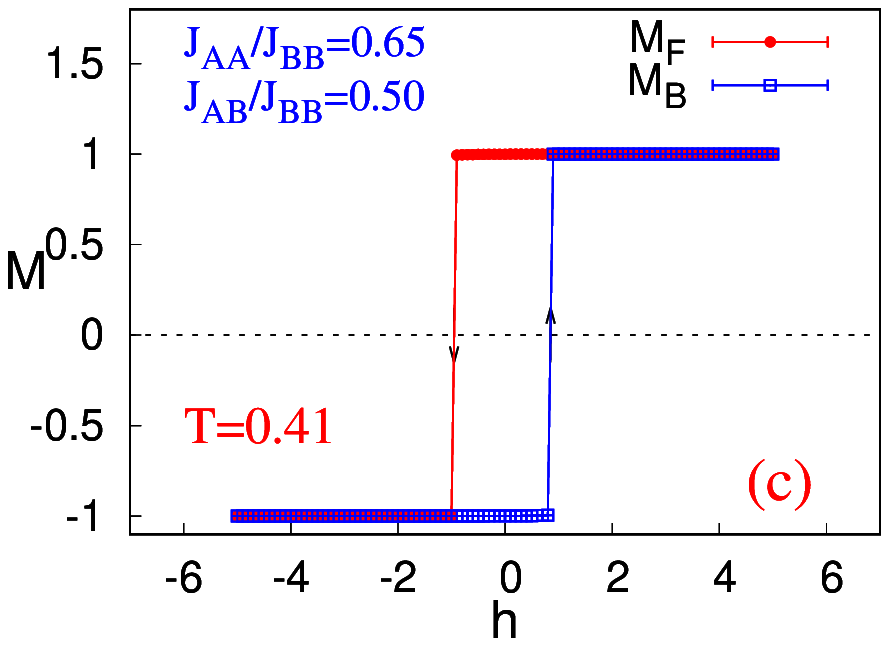}}
			\resizebox{7.5cm}{!}{\includegraphics[angle=0]{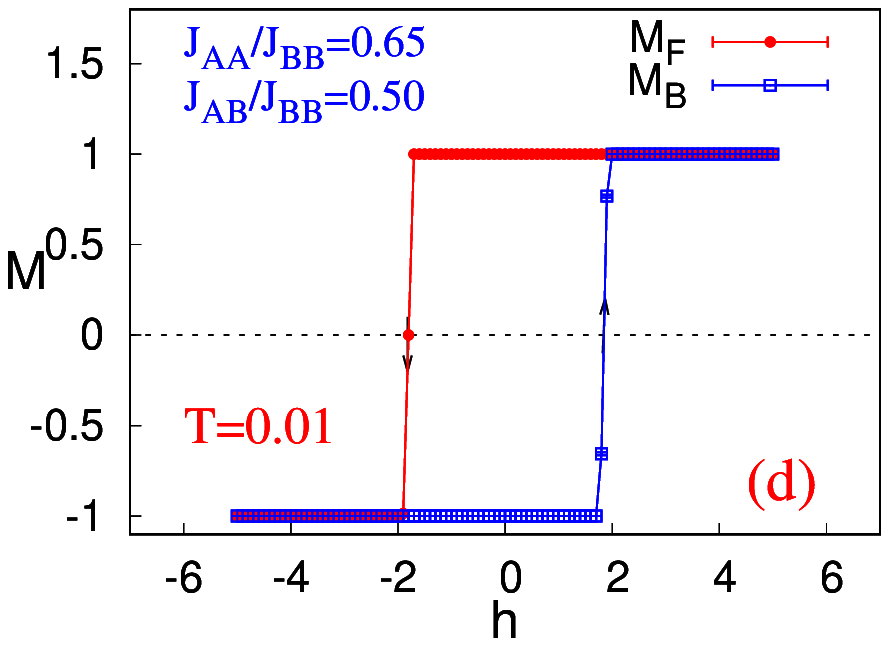}}
			
		\end{tabular}
		\caption{ (Colour Online) Plots of total magnetization versus the applied magnetic field. The effect of the temperature on the hysteresis behaviour for $T=1.31,0.81,0.41,0.01$ with the fixed values of $J_{AA}/J_{BB}=0.65$ and $J_{AA}/J_{BB}=0.50$. The descending/forward (RED, magnetisation $M_{F}$) and ascending/backward (BLUE, magnetisation $M_{B}$) external field directions are represented by arrows on each of the magnetization curves. As temperature of the system is gradually lowered, the area of the hysteresis loops increases.}
		\label{fig_3_hys_fr_fr}
	\end{center}
\end{figure*}

Here we have two scenarios: (a) $J_{AA}/J_{BB}=1.00$ and $J_{AB}/J_{BB}=0.50$ ; (b) $J_{AA}/J_{BB}=0.65$ and $J_{AB}/J_{BB}=0.50$. Because of the ferromagnetic interlayer coupling, the bulk would behave as a \textit{ferromagnet}, with different coupling strengths along the different crystallographic axes. We can see the effect of temperature on the shape of the hysteresis loops in Figures \ref{fig_2_hys_fr_fr} and \ref{fig_3_hys_fr_fr} respectively, where magnetization is plotted versus the external field with temperature acting as the parameter. All the four designated temperatures $(T={1.31,0.81,0.41,0.01})$ are lower than the critical point, $T_{c}$, in respective cases.

In both the cases, for all the temperatures we witness opening of a \textit{single, sharp, central} hysteresis loop. It is evident from the plots, as the temperature decreases, the area of the hysteresis loops increase. 

\subsection{Interlayer coupling: Antiferromagnetic}
\label{subsec_inter antiferro}

\begin{figure*}[!htb]
	\begin{center}
		\begin{tabular}{c}
			
			\resizebox{7.5cm}{!}{\includegraphics[angle=0]{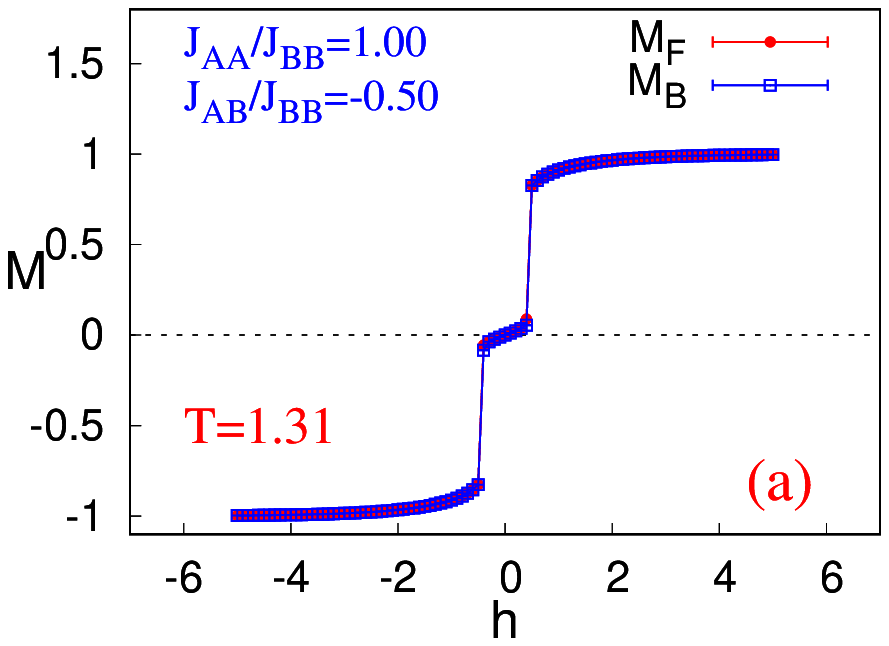}}
			\resizebox{7.5cm}{!}{\includegraphics[angle=0]{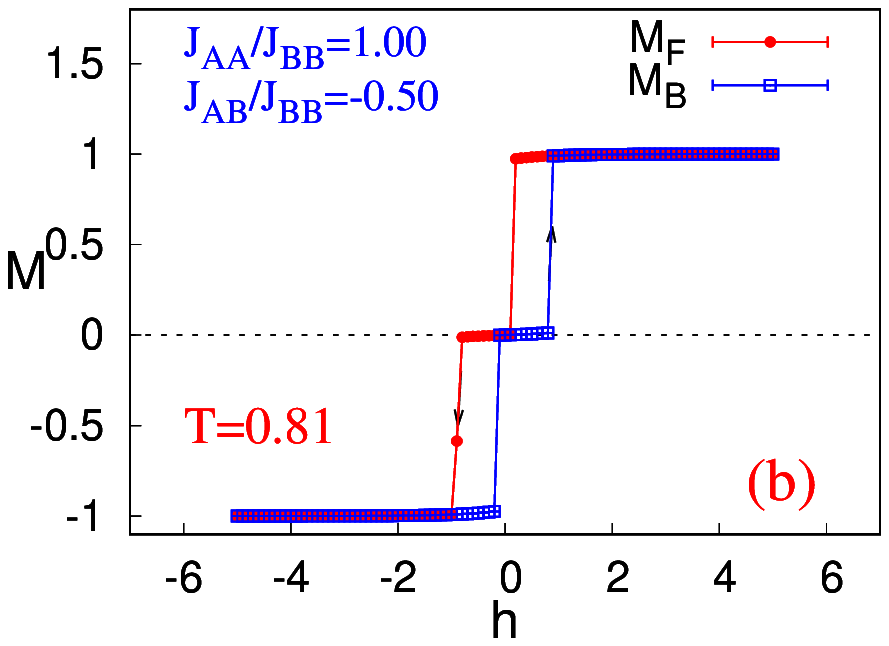}}\\
			
			\resizebox{7.5cm}{!}{\includegraphics[angle=0]{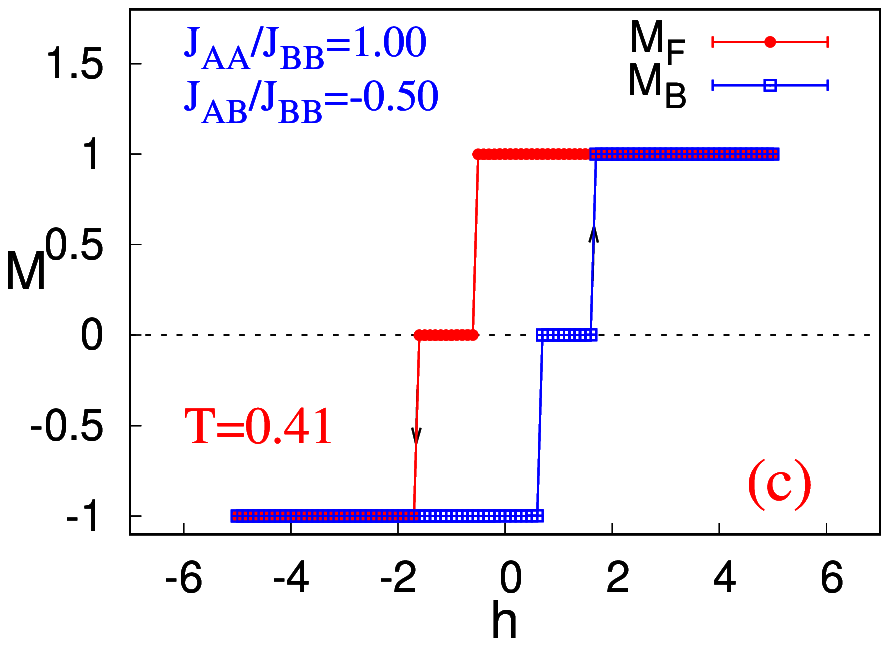}}
			\resizebox{7.5cm}{!}{\includegraphics[angle=0]{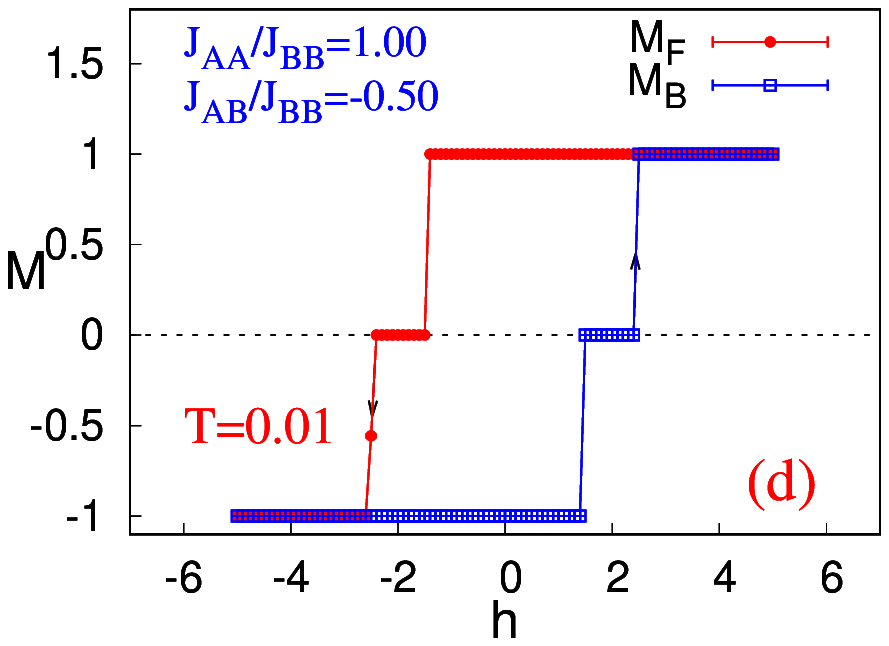}}
			
		\end{tabular}
		\caption{ (Colour Online) Plots of total magnetization versus the applied magnetic field. The effect of the temperature on the hysteresis behaviour for $T=1.31,0.81,0.41,0.01$ with the fixed values of $J_{AA}/J_{BB}=1.00$ and $J_{AA}/J_{BB}=-0.50$. The descending/forward (RED, magnetisation $M_{F}$) and ascending/backward (BLUE, magnetisation $M_{B}$) external field directions are represented by arrows on each of the magnetization curves. As temperature of the system is gradually lowered, the area of the hysteresis loops increases.}
		\label{fig_4_hys_fr_afr}
	\end{center}
\end{figure*}

\begin{figure*}[!htb]
	\begin{center}
		\begin{tabular}{c}
			
			\resizebox{7.5cm}{!}{\includegraphics[angle=0]{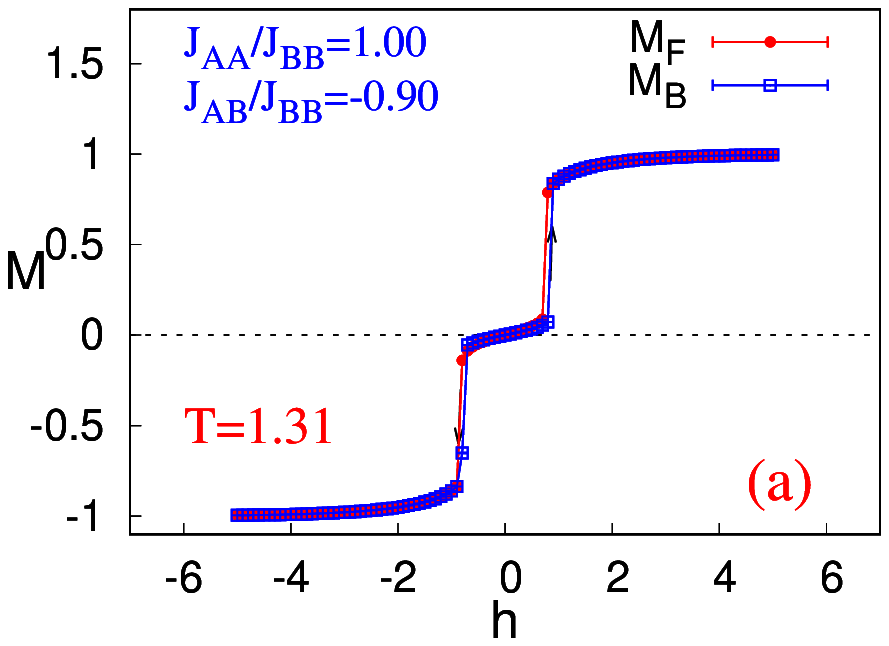}}
			\resizebox{7.5cm}{!}{\includegraphics[angle=0]{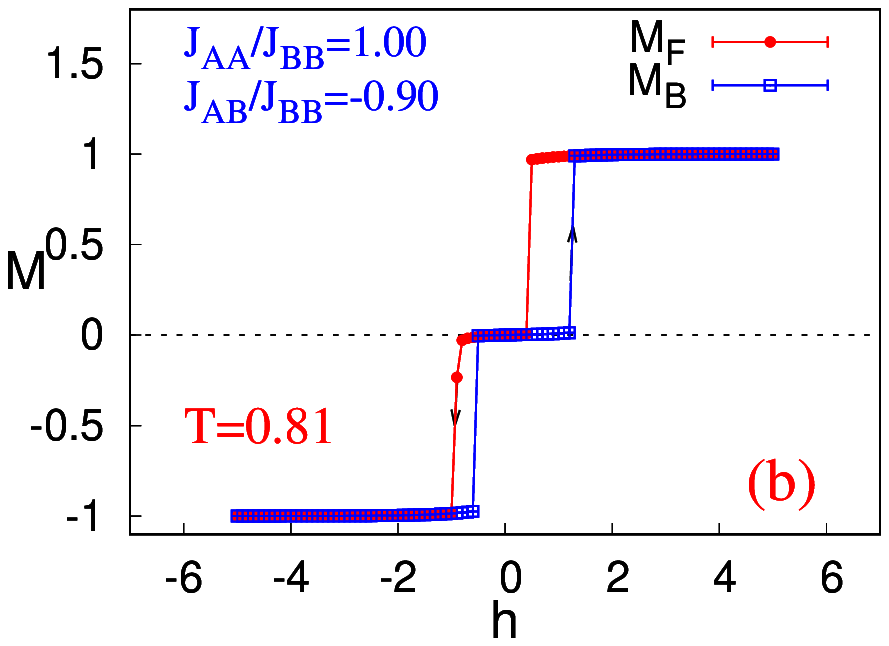}}\\
			
			\resizebox{7.5cm}{!}{\includegraphics[angle=0]{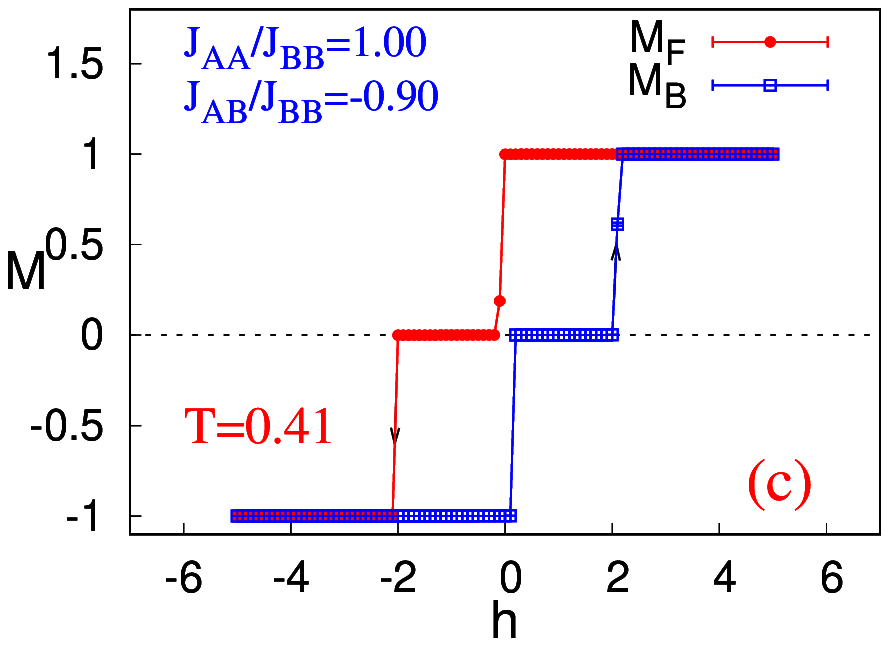}}
			\resizebox{7.5cm}{!}{\includegraphics[angle=0]{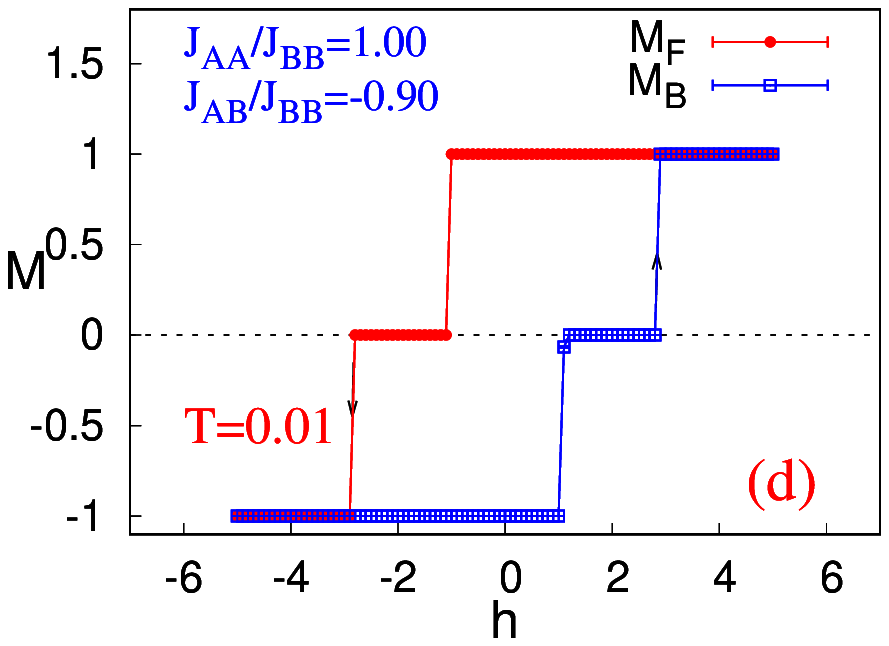}}
			
		\end{tabular}
		\caption{ (Colour Online) Plots of total magnetization versus the applied magnetic field. The effect of the temperature on the hysteresis behaviour for $T=1.31,0.81,0.41,0.01$ with the fixed values of $J_{AA}/J_{BB}=1.00$ and $J_{AA}/J_{BB}=-0.90$. The descending/forward (RED, magnetisation $M_{F}$) and ascending/backward (BLUE, magnetisation $M_{B}$) external field directions are represented by arrows on each of the magnetization curves. As temperature of the system is gradually lowered, the area of the hysteresis loops increases.}
		\label{fig_5_hys_fr_afr}
	\end{center}
\end{figure*}

We still keep, the intralayer coupling of the top A-layer ferromagnetic with $J_{AA}=J_{BB}$, but switch the interlayer coupling from ferromagnetic to antiferromagnetic. One combination is $J_{AA}/J_{BB}=1.00$ and $J_{AB}/J_{BB}=-0.50$ ; and the other combination is $J_{AA}/J_{BB}=1.00$ and $J_{AB}/J_{BB}=-0.90$. Because of the antiferromagnetic interlayer coupling and equal ferromagnetic intralayer coupling and spin value at both the layers, the bulk would behave as a perfect layered \textit{antiferromagnet}. The plots of magnetization versus the external field with temperature as a parameter in Figures \ref{fig_4_hys_fr_afr} and \ref{fig_5_hys_fr_afr} show the effect of temperature on the shape of the hysteresis loops. Again, all the four investigated temperatures are lower than the critical point, $T_{c}$, in respective cases.

We can see that changing the interlayer coupling from ferromagnetic to antiferromagnetic has a marked effect on the hysteresis loops of the system. For moderate antiferromagnetic interlayer coupling ($J_{AB}/J_{BB}=-0.50$), the hysteresis loop becomes a double loop with a plateau-like appearance about the zero field. The similar observation still holds good when the interlayer coupling is further decreased to make $J_{AB}/J_{BB}=-0.90$ .  

\section{Conclusion}
\label{sec_conclusion}

In this work, a Metropolis Monte Carlo study on the hysteretic behaviour of spin-1 bilayer Ising model on a square lattice is carried out. The effect of interlayer coupling on the hysteresis loops is indicated for a few temperatures below the critical points for the chosen cases. First, we found that in the ferromagnetic case, with $J_{AB}>0$, only a central hysteresis loop is available but as we shift to the antiferromagnetic case, with $J_{AB}<0$, the hysteresis loop changes to a double loop. So, we establish that besides a single hysteresis loop, double hysteresis loops can also appear in such a system. Second, as we decrease the temperature of the system, in both cases, the loop area increases. 

We could see that the values of magnetisation swept with the change in the external magnetic field are consistent with the available macroscopic magnetisation values, indicating that the magnetic interaction constraints are met. As a result, for antiferromagnetic interlayer coupling, we see a plateau-like region with $M=0$ (refer to Figures \ref{fig_4_hys_fr_afr} and \ref{fig_5_hys_fr_afr}), which is absent for ferromagnetic interlayer coupling. As the hysteresis loop is basically the variation of magnetisation with the external field at a fixed temperature, increasing (decreasing) the magnitude of spin associated with the magnetic atoms for the considered structure will lead to broadening (shrinking) of the loop along the vertical axis. Now for a fixed combination of the coupling strengths, as we lower the temperature, the layers tend to become more of a single domain according to the cooperative fields per atom. Thus, to make the macroscopic magnetic field vanish, we would need larger coercive field. This is responsible for the gradual growth of the area of the hysteresis loops with lowering of the temperature. As both the in-plane and interplane interactions lend magnetic stability (domain growth) to the magnetic heterostructure, increasing (decreasing) the magnitude of either of the coupling strengths will lead to higher (lower) coercive fields. This explains (and supported by the Figures \ref{fig_2_hys_fr_fr},\ref{fig_3_hys_fr_fr},\ref{fig_4_hys_fr_afr} and \ref{fig_5_hys_fr_afr}) the growth or shrink of the loops at a fixed temperature with changes in the combination of coupling strengths. 

Though extremely versatile, the well-known drawback of the effective field theory (EFT) with correlations is that it overestimates the critical temperature of a magnetic system. The same signature is observed in the study of \cite{Kantar} and can be compared with the values obtained in the current study. One way to provide a solution is by performing Monte Carlo simulation on such a system. A complete set of extensive simulations on the parameter space (with more \textit{equispaced} values of $J_{AA}$ and $J_{AB}$) may now be performed to completely characterize the hysteretic behaviour. Crystalline anisotropy is known to have an impact on the area of the hysteresis loops. This study also shows that the Metropolis Monte Carlo method can also be applied to similar kinds of studies in more complex situations, such as the mixed layered ferrimagnetic systems \cite{Chandra}. A fascinating extension of this current study is the response of this cooperative many-body magnetic system to an oscillating external magnetic field. Such a dynamic response in magnetic systems usually lags in time and creates a non-vanishing hysteresis loop. Two notable works to refer to in this direction, particularly on Ising systems, are \cite{Chakrabarti, Zhu}. These are planned for the future. 

\section*{Acknowledgements}
The author gratefully acknowledges the financial assistance from the University Grants Commission, India in the form of Research Fellowship.

\newpage
\section*{\begin{center}
		References
\end{center}}
\label{sec_references}

\end{document}